\documentclass[preprint,journal]{vgtc}       

\ifpdf
  \pdfoutput=1\relax                   
  \usepackage{graphicx}                
  \DeclareGraphicsExtensions{.pdf,.png,.jpg,.jpeg} 
\else
  \usepackage{graphicx}                
  \DeclareGraphicsExtensions{.eps}     
\fi%

\graphicspath{{figures/}{pictures/}{images/}{./}} 

\usepackage{microtype}                 
\PassOptionsToPackage{warn}{textcomp}  
\usepackage{textcomp}                  
\usepackage{mathptmx}                  
\usepackage{times}                     
\usepackage{cite}                      
\usepackage{tabu}                      
\usepackage{booktabs}                  

\usepackage{multicol}                  
\usepackage[dvipsnames]{xcolor}        
\usepackage{lettrine}                  
\usepackage{enumitem}
\setitemize{itemsep=1pt,topsep=2pt,parsep=0pt,partopsep=0pt}

\usepackage{url}

\PassOptionsToPackage{hyphens}{url}\usepackage{hyperref}

\usepackage[math]{cellspace}
\newcommand{\icon}[1]{\vspace{0.5em}\lettrine[lines=4, findent=2mm, nindent=-.5mm,image=true]{#1}{}}

\ieeedoi{xx.xxxx/TVCG.201x.xxxxxxx}

\onlineid{1312}

\vgtccategory{research}
\vgtcpapertype{theory/model}

\title{Table Scraps: An Actionable Framework for Multi-Table Data Wrangling From An Artifact Study of Computational Journalism}

\author{Stephen Kasica; Charles Berret; and Tamara Munzner, \textit{Senior Member}, \textit{IEEE}}
\authorfooter{
\item
 Stephen Kasica and Tamara Munzner are with the University of British Columbia, Department of Computer Science. E-mail: \{kasica,tmm\}@cs.ubc.ca.
\item
 Charles Berret is with the University of British Columbia, School of Journalism, Writing, and Media. E-mail: cberret@mail.ubc.ca.
}

\shortauthortitle{Kasica \MakeLowercase{\textit{et al.}}: Table Scraps}

\abstract{For the many journalists who use data and computation to report the news, data wrangling is an integral part of their work. Despite an abundance of literature on data wrangling in the context of enterprise data analysis, little is known about the specific operations, processes, and pain points journalists encounter while performing this tedious, time-consuming task. To better understand the needs of this user group, we conduct a technical observation study of 50 public repositories of data and analysis code authored by 33 professional journalists at 26 news organizations. We develop two detailed and cross-cutting taxonomies of data wrangling in computational journalism, for actions and for processes. We observe the extensive use of multiple tables, a notable gap in previous wrangling analyses. We develop a concise, actionable framework for general multi-table data wrangling that includes wrangling operations documented in our taxonomy that are without clear parallels in other work. This framework, the first to incorporate tables as first-class objects, will support future interactive wrangling tools for both computational journalism and general-purpose use. We assess the generative and descriptive power of our framework through discussion of its relationship to our set of taxonomies.} 


\begin{document}


\firstsection{Introduction}

\maketitle

Data wrangling is an exploratory, iterative process of auditing and transforming data, encompassing tasks such as cleaning, integrating, and transforming datasets as an often necessary precursor for data analysis~\cite{kandel_wrangler_2011}. It is also an arduous process comprising a significant portion of effort data analysis and data warehousing projects~\cite{muller_how_2019,dasu_exploratory_2003,kandel_enterprise_2012}. Programmatic wrangling is typically carried out with general-purpose scripting languages such as R or Python, often augmented with supplemental libraries~\cite{wickham_dplyr_2019,wickham_tidyr_2019}. In addition, several interactive tools have been designed to support data wrangling among data-literate non-programmers~\cite{kandel_wrangler_2011, bigelow_origraph_2019, noauthor_trifacta_2012, huynh_openrefine_2012, stray_workbench_2018}, incorporating visualization to assist in data auditing and evaluation of data transformations.

Observational studies of the data wrangling process could guide the design and evaluation of wrangling support systems, both programmatic and interactive. The interview study of enterprise data analysts~\cite{kandel_enterprise_2012} was a useful start, but many questions remain open. We choose to study a specific domain that is an microcosm of data wrangling: computational journalism.  Journalists have two amenable characteristics as a target population: a clear need to wrangle and a culture of extensive process documentation. First, the need for transforming and cleaning raw data has been identified as a preeminent challenge~\cite{welsh_what_2018}. Second, the culture of journalism valorizes transparency and providing evidence for reported conclusions~\cite{anderson_apostles_2018,schudson_discovering_1978}. Many data-driven articles conclude with a methodology statement, informally known as the ``nerd box,'' that typically includes a link to publicly released code to replicate the entire underlying analysis process, including all wrangling operations.

We conduct a technical observation study~\cite{ralph_toward_2019} of how professional journalists use scripting languages to wrangle data in the wild, from code repositories (repos) that journalists have made publicly available in conjunction with published articles. We study the actions of journalists who program as a first step in understanding the essential operations that should be supported for both programmers and non-programmers.  By studying the data wrangling actions performed with the power and flexibility of code, we can better understand what an interactive data wrangling interface should provide to non-coding journalists.

Our work at the intersection of computer science and journalism is descriptive of computational journalism but not exclusive to it. Journalists created the artifacts used in our study with common data science tools, and we find that they encounter issues similar to users in other domains, such as enterprise data analysis~\cite{kandel_research_2011}.

Our observational study results in two descriptive taxonomies of data wrangling in computational journalism. These taxonomies, grounded in data, are created by the bottom-up method of open and axial coding on the technical artifacts of programming scripts and computational notebooks from a set of repos that we curated. One taxonomy describes the actions journalists took while wrangling their data, and the other features our interpretation of their wrangling processes. 

The most novel finding of our study is the extent to which journalists make use of multiple tables in their wrangling activities, in contrast to previous wrangling frameworks that emphasize wrangling operations conducted within a single table. To fill this gap, we present an actionable framework for multi-table wrangling. The framework is designed with a concise structure that provides generative power to serve as the basis for building future interactive tools. We synthesize the framework through a top-down approach where we consider tables themselves as first-class objects, with equal footing to its rows and columns. We cross-check the coverage of the framework by ensuring it covers the many multi-table operations observed in our action taxonomy that do not fit in existing frameworks for data wrangling and harmonize with concepts and vocabulary from the existing literature. 

We present two primary contributions: two detailed descriptive taxonomies of data wrangling in computational journalism and a concise framework for multi-table data wrangling designed to be an actionable basis for developing future general-purpose tools.

We also present two secondary contributions of document corpora as supplemental material. The first corpus provides links to the curated set of 225 repos. The second corpus is the subset of 50 repos used in our study, fully annotated with open codes to provide transparent supporting evidence of our analysis process~\cite{meyer_criteria_2020}. These secondary contributions may also prove useful for researchers interested in better transformation recommendations in data wrangling~\cite{kandel_research_2011}.

\section{Related Work}

We review related work on characterizing the data wrangling process in journalism and computer science. 

\subsection{Data Wrangling in Journalism}

A small body of work characterizes wrangling challenges faced by journalists, often in the form of retrospective articles about the data analysis process for individual stories as well as processes used by the news organization in general. Nguyen~\cite{nguyen_dollars_2010} details how a team of journalists at \textit{ProPublica} used Google Refine to perform entity resolution on data underlying an investigation into payments to US doctors from pharmaceutical and medical device companies. Rogers~\cite{rogers_data_2013} outlines a general data journalism workflow where common data cleaning tasks constitute early steps, estimating that data journalists spend the majority of their time engaged in cleaning and merging datasets~\cite{rogers_data_2011}. The \textit{Quartz} Bad Data Guide~\cite{groskopf_quartz_2015} enumerates many real-world data issues faced by journalists. While these works are largely issue focused, concentrating on common data errors and issues, our descriptive taxonomy is action focused, concentrating on ways to resolve data issues.

Although the majority of material on wrangling in journalism comes via popular articles, there have been a few academic and industry studies. Cohen et al.~\cite{cohen_computational_2011} asserts that cleaning and verifying the contents to merge multiple data sources is a common task in data journalism. A recent survey of data journalists found that about half of respondents reported creating data-driven stories in a day or less, including time spent wrangling~\cite{rogers_data_2017}. A report from The American Press Institute characterizes the skill levels for common wrangling operations supported by spreadsheet applications~\cite{sunne_diving_2016}.

Although many journalists use general-purpose wrangling tools, the tools they build for themselves also provide insight into issues they face. The command-line tool \texttt{csvkit}~\cite{groskopf_csvkit_2012} and its Python equivalent \texttt{Agate}~\cite{groskopf_agate_2018} provide functionality for wrangling tasks such as unique key generation, pivoting a table by one or more columns, and deriving additional columns. Workbench~\cite{workbench_data_2018} is a data journalism platform built around community-contributed modules for wrangling, analysis, and visualization. While the process of eliciting design requirements for these tools has not been clearly documented, our work provides a systematic characterization to inform future tool design in this area.

\subsection{Data Wrangling in Computer Science}

Tools for wrangling data fall into two categories: scripts written in various programming languages and interactive applications.

Scripting languages such as Python and R have commonly been used to wrangle data~\cite{kandel_wrangler_2011}, often with supplemental libraries to further facilitate the process. Pandas~\cite{reback_pandas-devpandas_2020}, plyr~\cite{wickham_split-apply-combine_2011}, dplyr~\cite{wickham_dplyr_2019}, and tidyr~\cite{wickham_tidyr_2019} are widely used for data wrangling in a programming environment. These packages incorporate structures for representing heterogeneous data within their environment as analogous to tables. Both Python and R have tools to instantiate the wrangling design principles of tidy data~\cite{wickham_tidy_2014} and the split-apply-combine strategy for data analysis~\cite{wickham_split-apply-combine_2011}. Our work synthesizes and incorporates these ideas in our multi-table framework in a way that can be applied to interactive interfaces.

While general-purpose tools like Microsoft Excel implicitly support many common wrangling operations, several interactive tools designed explicitly for wrangling offer an expanded set of operations accessible to empower data-literate non-programmers. OpenRefine~\cite{huynh_openrefine_2012}, Trifacta Wrangler~\cite{noauthor_trifacta_2012}, and Tableau Prep Builder~\cite{noauthor_tableau_2019} enable users to clean, edit, and merge data using a menu-based GUI and application-specific custom functions. These systems leverage data visualization in an iterative, human-in-the-loop process of data auditing and transformation. Trifacta Wrangler (also known as Google Cloud Dataprep) and its predecessor Wrangler~\cite{kandel_wrangler_2011} incorporate transformation recommendations throughout the wrangling process. These applications invoke a spreadsheet or matrix view where the table is synonymous with the environment. While they may support operations to merge multiple data sources, they do not do so by treating the table as a first-class citizen in the wrangling environment.

While many existing wrangling applications do support some operations that involve multiple tables, none support the whole breadth of possible operations within this space. Data integration is a sub-problem within wrangling~\cite{kandel_wrangler_2011} and mashup applications are often cited as examples of interfaces that address this aspect of wrangling~\cite{kandel_research_2011, kandel_wrangler_2011}. However, these applications are primarily concerned with extracting data from HTML web pages and secondarily incorporating extracted data into other structured data. Many of them support JOIN-like operations~\cite{wong_making_2007, tuchinda_building_2008} common in Structured Query Language (SQL) or perform data integration without the notion of a table~\cite{huynh_potluck_2008,lin_end-user_2009}. None of these applications support the kind of operations that map one table to many tables like we observed users performing in our technical observation study. 

Some research addresses wrangling network data. However, we did not observe journalists extensively performing network analysis techniques, and previous work addresses just how rarely network analysis is used by journalists~\cite{stray_network_2017}. Ploceus~\cite{liu_network-based_2011} and Orion~\cite{heer_orion_2011} implicitly support basic graph editing operations. Origraph~\cite{bigelow_origraph_2019} explicitly supports network wrangling, expanding upon these wrangling operations. Concepts from these papers inspire some of the operations in our multi-table framework; however, these systems do not directly support multi-table wrangling of tabular data.

There are only a few observational studies of the data analysis process that includes data wrangling. Kandel et al.~\cite{kandel_enterprise_2012} performed an interview study of enterprise data analysts. Muller et al.~\cite{muller_how_2019} perform an interview study with data scientists, and describe five ways data science workers engage with data. Our work differs in both domain and methodology. We focus on wrangling by computational journalists using programming languages, and we use observation of technical artifacts as our data collection method.

\begin{figure*}
    \centering
    \includegraphics[width=\linewidth]{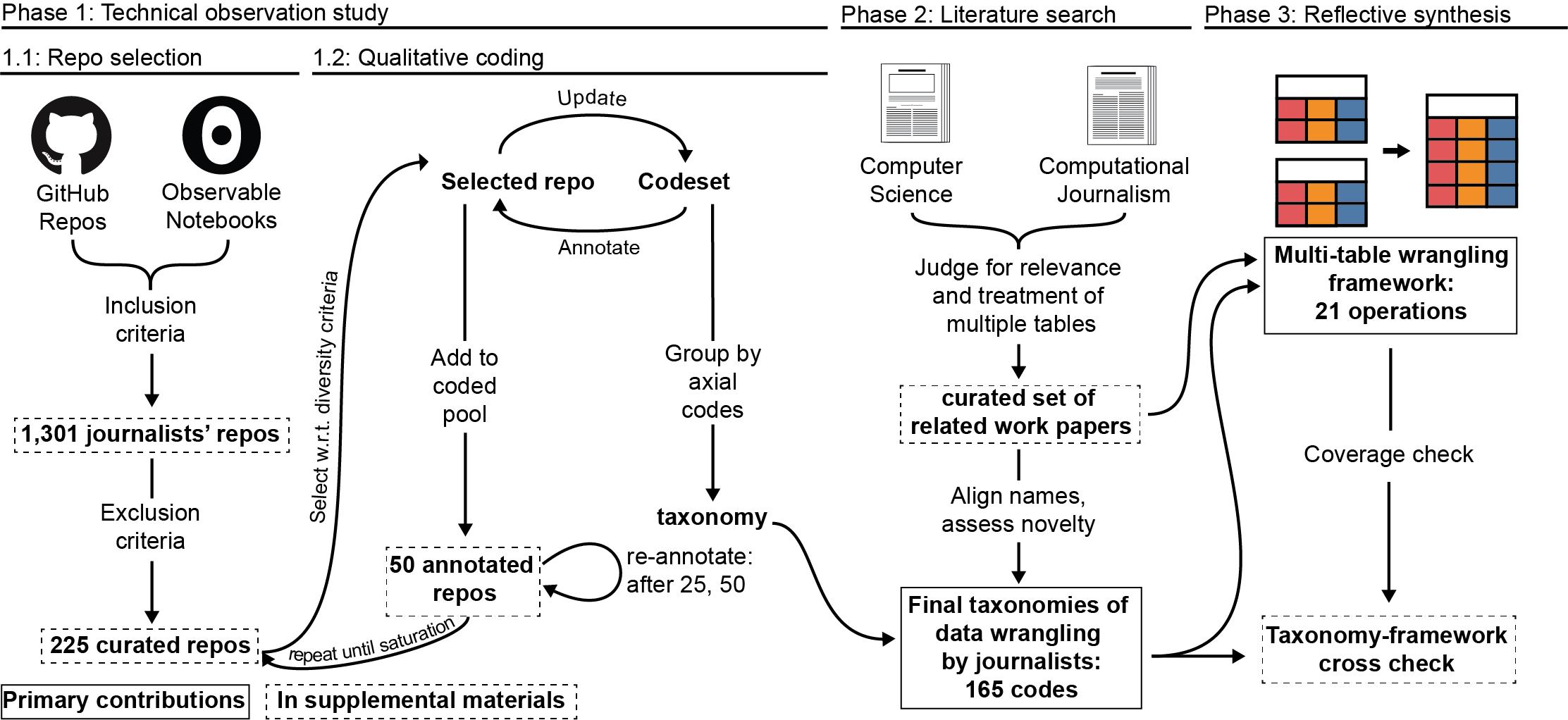}
    \caption{Three-phase process: observation study of technical artifacts conducted through qualitative coding of journalist repos, resulting in two initial bottom-up taxonomies of 165 open and axial codes; literature search to align naming and assess novelty; reflective synthesis to create a concise top-down multi-table wrangling framework with 21 operations.}
    \label{fig:process}
    \vspace{-1em}
\end{figure*}

\section{Process and Methods}

The research questions addressed in this paper are:
\begin{enumerate}[label=Q\arabic*:,itemsep=0em]
\item \textit{What are the wrangling practices of data-literate journalists with programming skills?}
\item \textit{Which practices align with or diverge from existing characterizations?}
\item \textit{How to re-characterize wrangling to match the observed practices?}
\end{enumerate}

Our process has three phases, one for each question (Figure \ref{fig:process}). The first phase studies the wrangling processes of journalists using the qualitative method of open and axial coding on technical artifacts. These artifacts document the analysis underpinning articles and are published in conjunction with them in publicly available repositories. The product of this phase is an initial bottom-up, descriptive taxonomy of data wrangling in computational journalism. The second phase employs an interdisciplinary literature search to compare our taxonomy to the existing literature on process theories of data wrangling. We conduct The third phase via reflective synthesis, to create a concise framework for multi-table wrangling with 21 operations. We then check that this cross-cutting framework fully covers all of the wrangling actions we observe in the study and document in the bottom-up taxonomy.

\subsection{Phase One: Qualitative Coding Study Overview}

In the first phase we addressed Q1 through qualitative coding of programming scripts and computational notebooks supporting published articles. We systematically searched GitHub and ObservableHQ to identify an initial corpus of more than 1,000 journalistic code repositories related to journalism. From these initial results, we inspected each repository to identify those containing data analysis, resulting in a set of 225 repositories.

This curated corpus served as the basis for the data in our technical observation study. Through an iterative process of manually selecting repos according to criteria that ensure diversity of both individuals and organizations, we produced a final set of 50 annotated repos documenting journalists' data analysis process with open codes. Through axial coding, we produce a descriptive, bottom-up taxonomy of wrangling in computational journalism grounded in this observational data.

Studies of provenance in E-Science make a distinction between whether records are data or process oriented~\cite{simmhan_survey_2005, ragan_characterizing_2016, simmhan_performance_2006}. We also distinguish between data and process in the qualitative coding portion of the first phase. The cross cutting nature of our taxonomy occurs along two dimensions: wrangling \textit{actions} performed by the journalists upon the data, which are orthogonal to descriptions of the wrangling \textit{process}.

In related work on data wrangling, journalists are considered a non-technical user group~\cite{kandel_wrangler_2011}, but there is a data-literate subset of journalists who routinely practice technical tasks such as data wrangling, database management, statistical analysis, and data visualization. These data-literate journalists can be further classified by how much of this work is performed with programming languages versus interactive tools. The repo authors in this corpus are a sample of technically savvy data journalists fluent in various programming languages. Although this group does constitute the minority of data journalists~\cite{sunne_diving_2016, global_investigative_journalism_network_nils_nodate}, we expect theories generated from studying the practices of data-literate programmer journalists will transfer to the larger population of data-literate non-programmer journalists. We conjectured that these programmatic approaches to wrangling would be more expressive than the operations supported by interactive applications, which aim to democratize the tools and techniques for data manipulation~\cite{kandel_wrangler_2011, bigelow_origraph_2019}.

\subsubsection{Observation of Technical Artifacts}

We employ the term \textit{technical observation study} from the software engineering literature to denote a data collection strategy of observing user-generated technical artifacts, such as source code~\cite{ralph_toward_2019}. This approach is similar to indirect observation, as both methods involve mediated, \textit{post hoc} analysis of a user group. Whereas indirect observation studies collect data through researcher-developed tools like keystroke logs or transcriptions of audio and video recordings~\cite{sharp_interaction_2019}, technical observation data is an artifact of the phenomenon itself.

Wrangling is especially well suited to this technical-observational approach. The chief product of wrangling is not only the transformed and cleaned data, but also the record of the transformations applied to the raw data~\cite{kandel_wrangler_2011, kandel_research_2011}. Programming scripts and computational notebooks constitute an auditable and reproducible transformation record. As a result, this approach is positioned to produce theories of \textit{how} users wrangle their data with strong ecological validity. Thus, when forming our taxonomy, we implemented this bottom-up approach of qualitative coding by annotating journalists' scripts and notebooks with open and axial codes, grounding our findings in these transformation records.

Technical observation allows us to quickly and easily analyze data with high ecological validity with no demands on the target population's time. This approach does have limitations. We limit our claims to focus on \textit{how} journalists wrangle data, with limited conjecture into \textit{why} they perform these actions. While code comments and the before-and-after state of a table may provide some sense of the user's motivation, qualitative research methods such as interviews and direct observation are better suited to this question~\cite{sharp_interaction_2019}. Although previous work notes that those engaged in data analysis often explore alternatives~\cite{liu_understanding_2020}, these repos typically contain a straightforward pipeline from the raw data to its final form, and may omit false starts and dead ends. Moreover, our collection process filters out all instances of unsuccessful wrangling. Our study may thus paint a simpler view of wrangling than the reality.

\subsubsection{Repository Selection}

We compiled an initial corpus of public code repositories from GitHub and ObservableHQ by two inclusion criteria: being relevant to journalism and being written in a common programming language used for data wrangling. We use the term \textit{repo} to refer to any collection of related materials in either platform. This process concluded with a set of curated repos containing journalists data analysis, which we include in supplemental materials.

On GitHub, we identified journalistic repos through two avenues. We conducted a programmatic search using the platform's Search API, parameterized by topic, owner, and programming language. We satisfied the relevance to journalism criterion by identifying repos with a \textit{journalism} or \textit{data journalism} topic tag. We also referenced \texttt{@NewsNerdRepos}, a Twitter bot that posts new repos published by journalists. Any GitHub user or organization monitored by this account\footnote{github.com/silva-shih/open-journalism} satisfied the journalism-relevance criterion. We satisfied the wrangling language criterion by restricting our search to those repos where the predominant programming language in the repo is R, Python, or Jupyter Notebook. We chose R and Python due to their inclusion in previous wrangling papers~\cite{bigelow_origraph_2019, kandel_wrangler_2011} and added Jupyter due to its rising popularity for data analysis. In order to gather R Markdown files, an idiosyncrasy of GitHub forced us to include HTML in the search parameters, leading to the inclusion of many web applications such as front-end visualizations or news applications that do not contain examples of data wrangling. We manually inspected the contents of each repo to exclude irrelevant ones, yielding a much smaller corpus of curated repos.

ObservableHQ is a computational notebook environment similar to Jupyter, where Observable notebooks are created using the front-end web development tools HTML, CSS, and JavaScript. We included a total of 36 repos from three journalists at major daily newspapers who were panel members of a conference workshop on this topic~\cite{chinoy_first_2019}, automatically satisfying the relevance criterion. The portion of these repos applicable to wrangling are written in JavaScript, adding the diversity of another wrangling language to our corpus of coded repos. After similar manual filtering, we retained 34 repos from this set.

\subsubsection{Qualitative Coding}

We select a single repo at a time from the curated corpora to analyze through a manual selection procedure using two inclusion criteria: an explicit connection to a published article and increased diversity in experience or practices. By deliberately selecting only repos with corresponding articles, we ensure that our results are based on work where journalists were successful in wrangling raw data for their analysis or visualization needs. We deliberately select repos authored by different journalists at different news organizations to ensure a wide range of tools, practices, and experience. With each repo, we add new codes to our codeset, refine existing codes, and opportunistically apply new codes to previous notebooks. We performed axial coding in batches, splitting and consolidating code groups as needed. The intent of previously encountered behavior often became clearer as the coder encountered similar examples. In addition to opportunistic, retroactive application of codes, the coder also systematically re-coded earlier work after completing 25 and 50 repos, at which point we reached theoretical saturation, concluding the first phase.

Once collecting more data resulted in diminishing returns with respect to information that influenced our newly constructed theories, we determined that our study had reached saturation and required no further data. Two factors signaled reaching this point. First, the cardinality of our codeset began to level off, and the internal structure ceased to change. Adding new codes did little to modify the internal structure of axial codes. Thus, our taxonomies adequately described new phenomena. Second, the first author drew upon previous experience as a journalist and assessed that our theory conveyed a thorough understanding of major themes; sustained engagement in a field can help researchers achieve theoretical saturation~\cite{given_sage_2008}. At the conclusion of this stage, the open and axial codes were exported as the initial taxonomies.

All repos from the curated corpora were analyzed by a single coder, the first author. We use established qualitative criteria: transparent reporting of our procedure, thick description of sample data, and triangulation with related research. In contrast, quantitative analysis with a closed codeset often involves a positivist approach with multiple coders who converge above a threshold for inter-coder reliability to demonstrate replicability~\cite{sharp_interaction_2019}, but we felt the interpretive agenda of qualitative research was a better match with our goals.

While coding the first initial notebooks, we began with a single taxonomy, but further into this process, we saw two different types of code emerging, regardless of the level at which they describe phenomena. The first comprises \textit{actions} journalists made upon their data in which we could reasonably infer their motivation based on the consequence of their operations and the semantic APIs of certain coding-based wrangling tools. The second concerned our own observations about the wrangling \textit{process} at a level higher than just short sequences of transformation operations. Thus, we retroactively pivoted one taxonomy into two separate taxonomies, one for \textit{actions} and one for \textit{processes}.

\begin{figure}[h!]
    \centering
    \includegraphics[width=8cm]{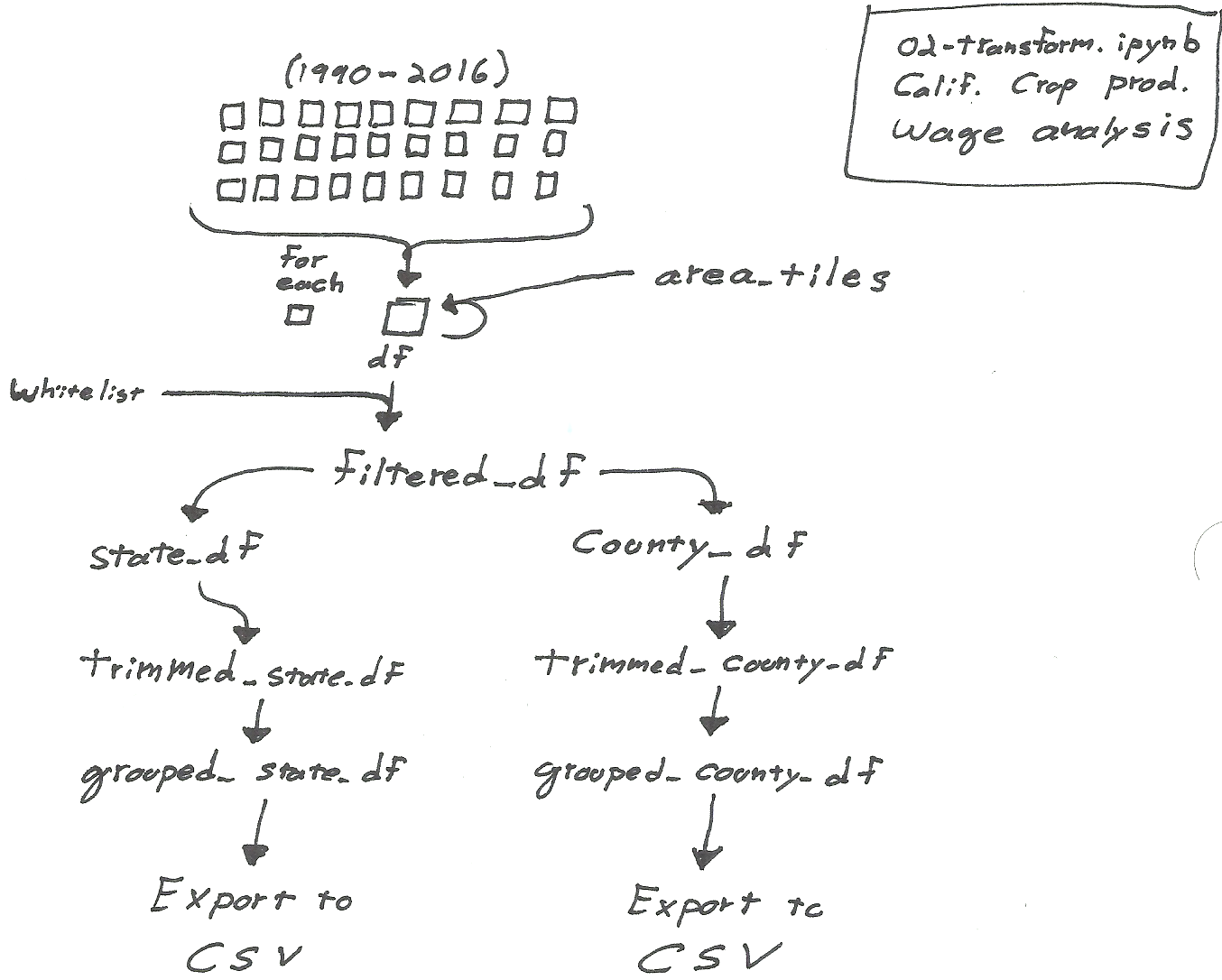}
    \caption{A sketch of data flow through a notebook authored by journalists at the Los Angeles Times shows a wrangling process using more than two dozen tables before exporting two datasets for analysis and visualization.}
    \label{fig:table-graphs}
    \vspace{-1em}
\end{figure}

\textbf{Data Flow Sketches:} We quickly noticed that journalists frequently employ multiple tables. To facilitate our own understanding of their activity, we sketched table-based data flow diagrams of how raw data is transformed through the wrangling environment when tables are used in complex ways. Figure~\ref{fig:table-graphs} shows an example. These diagrams were instrumental for the central finding, reflected in our taxonomies, that journalists often employ many tables in ways not addressed by previous characterizations of wrangling operations.

\subsection{Phase Two: Literature Review}

In the second phase, we address Q2. Through a search of relevant literature, our goals are to reconcile the labels of our codes with terminology from the research literature and to assess the novelty of phenomena observed in the previous phase.

We conduct our literature search in two stages, both performed by the first author. The first stage involves assembling a set of seed papers from two research domains: data wrangling in the computer science literature and journalistic data analysis in both popular and academic literature. We identify seed papers through targeted searching informed by the authors' background knowledge. Following the creation of preliminary observational taxonomies, we expand this set by following citations of papers that address the role of multiple tables and data sources in the wrangling process, using these papers' discussion of data integration, mashups, or network wrangling as inclusion criteria. We review this expanded set of seed papers, noting those with similarities and differences to observed phenomena. We harmonize the nomenclature in our taxonomy to align with previous usage to create the final version of the taxonomy, resulting in dozens of name changes. For example, \textit{Create a Semi-Unique Key} was an early code describing users generating keys with no guarantee or assessment of uniqueness, such as concatenating given names and family names. We rename this code to \textit{Create Soft Key} to align with literature in data cleaning~\cite{dasu_exploratory_2003}.

\subsection{Phase Three: Reflective Synthesis}

In the final phase, we address Q3. By reflecting on other taxonomies of data wrangling operations~\cite{kandel_wrangler_2011,bigelow_origraph_2019,heer_orion_2011} in relation to our observational findings, we synthesize a design space for multi-table data wrangling. Many interactive wrangling applications specify a design space through a domain specific language or enumerate supported operations. While these design spaces are informed by work in data transformation languages or personal experience in data wrangling, our work is grounded in observational data gathered from users wrangling data in the wild with the full flexibility of scripting languages. The result of this process is our Framework for Multi-Table Wrangling, described in Section~\ref{sec:framework}.

The motivation for this second framework is our interest in finding a process model to guide the design of existing and future wrangling tools. The utility of a conceptual framework for guiding designers can be evaluated in terms of both descriptive and generative power~\cite{beaudouin-lafon_designing_2004}. Our set of descriptive taxonomies is centered on users, organized around actions, and observed in wrangling processes. This set of taxonomies provides a rich vocabulary for describing the actions and processes of journalists. However, they lack generative power: it would be very difficult for a designer to use them for guidance in building a system.

In contrast to the user-centered focus of our taxonomies, this framework centers on the state of the data itself, defining transformation operations by the cardinality and type of its inputs and outputs. One hurdle when navigating the design space of wrangling transformations is the inconsistent nomenclature used across computational, statistical, database, and data wrangling literature. A data-centered categorization of wrangling operations abstracts away this terminology and will help designers build better tools.

\section{Descriptive Taxonomies: Actions and Processes}

Our bottom-up taxonomies of data wrangling in computational journalism provide richly detailed descriptions of the \textit{Actions (A)} journalists took while wrangling their data and our interpretation of their wrangling \textit{Processes (P)}. Table \ref{tbl:taxonomy-table} provides an overview of the whole taxonomy down to a depth of two. Unabridged versions of these taxonomies are provided in Supplemental Materials, with shortcode, name, and description for each code. Below, each code name is followed by the shortcode in parentheses to facilitate its lookup.

These two taxonomies thematically organize 165 open and axial codes. Both taxonomies are hierarchically organized into wide trees with a maximum depth of three. The \textit{Actions (A)} taxonomies and \textit{Process (P)} taxonomies contain 66 open codes with 25 axial codes and 56 open codes with 18 axial codes, respectively. The tree is root-justified not leaf-justified: open-code leaves may occur at any level.

\subsection{High-Level Overview}
Our taxonomy exists in two orthogonal and cross-cutting dimensions: wrangling \textit{Actions (A)} performed on the data and descriptions of the wrangling \textit{Process (P)}.

\begin{table}[h!]
\label{tbl:taxonomy-table}
\begin{tabular}{| m{4cm} | m{4cm}| }
\hline
\textbf{Actions} & \textbf{Process} \\ 
\hline
\hspace{0mm} \textbf{Import} & \hspace{0mm} \textbf{Source} \\ 
\hspace{2mm} Fetch & \hspace{2mm} Collect Data \\ 
\hspace{2mm} Create & \hspace{2mm} Acquire Data \\ 
\hspace{2mm} Load & \hspace{0mm} \textbf{Workflow} \\ 
\hspace{0mm} \textbf{Clean} & \hspace{2mm} Annotations \\ 
\hspace{2mm} Remove & \hspace{2mm} Comp. Processes \\ 
\hspace{2mm} Replace & \hspace{2mm} Toggle Operation \\
\hspace{4mm} Replace NA Values & \hspace{0mm} \textbf{Cause} \\ 
\hspace{4mm} Edit Values  & \hspace{2mm} Downstream Input \\ 
\hspace{4mm} Resolve Entities & \hspace{0mm} \textbf{Themes} \\  
\hspace{4mm} Standardize Cat Vars & \hspace{2mm} Divide and Conquer \\ 
\hspace{4mm} Scale Values & \hspace{2mm} Join Aggregate \\ 
\hspace{2mm} Reformat  & \hspace{2mm} Create a Frequency Table \\ 
\hspace{0mm} \textbf{Merge} & \hspace{2mm} Trim Fat \\ 
\hspace{2mm} Union Datasets & \hspace{2mm} Align Variables \\  
\hspace{2mm} Inner Join  & \hspace{0mm} \textbf{Analysis} \\ 
\hspace{2mm} Supplement  & \hspace{2mm} Interpret Model \\ 
\hspace{2mm} Cartesian Product  & \hspace{2mm} Compare Groups \\ 
\hspace{2mm} Self Join Dataset & \hspace{2mm} Identify Extreme Values \\ 
\hspace{0mm} \textbf{Profile} & \hspace{2mm} Show Trend Over Time \\ 
\hspace{2mm} Run a Test & \hspace{2mm} Calculate a Statistic \\ 
\hspace{2mm} Check Results  & \hspace{2mm} Count the Data \\
\hspace{2mm} Summarize Dataset  & \hspace{2mm} ... \\
\hspace{0mm} \textbf{Derive}  & \hspace{0mm} \textbf{Management} \\ 
\hspace{2mm} Detrend & \hspace{2mm} Object Persistence \\ 
\hspace{2mm} Consolidate Variable Values  & \hspace{2mm} Data Quality \\ 
\hspace{2mm} Generate Unique Identifiers & \hspace{0mm} \textbf{Pain Points} \\ 
\hspace{2mm} Subset the Dataset & \hspace{2mm} Fix Incorrect Calculation \\ 
\hspace{2mm} Formulate Perf Metric  & \hspace{2mm} Repetitive Code \\ 
\hspace{0mm} \textbf{Transform} & \hspace{2mm} Make an Incorrect Conclusion \\ 
\hspace{2mm} Reshape & \hspace{2mm} Post-Merge Clean Up \\ 
\hspace{2mm} Modify Variables & \hspace{2mm} Post-Aggregate Clean Up \\ 
\hspace{2mm} Summarize & \hspace{2mm} Data Too Large for Repo \\ 
\hspace{2mm} Sort & \hspace{2mm} Schema Drift \\ 
\hspace{0mm} \textbf{Export}  & \hspace{2mm} Data Type Shyness \\
\hline
\end{tabular}
\smallskip
\caption{Abridged version of our two descriptive taxonomies of data wrangling in computational journalism. The unabridged taxonomies extend to five levels, and are provided as supplemental materials.}
\vspace{-2em}
\end{table}

\textbf{Actions}: We record seven high-level groups within the actions taxonomy. \textit{Import (A.I)} captures how data is introduced to the wrangling environment. The \textit{Clean (A.C)} branch addresses actions for well-known data quality issues, such as entity resolution, deduplication, and addressing missing or incomplete data. We record operations that combine multiple tables together under \textit{Merge (A.M)}. Journalists often inspect the state of a table either before or after a transformation under the category \textit{Profile (A.P)}. The \textit{Derive (A.D)} branch contains codes for operations that transform observations and variables within a dataset, but do not necessary address data quality issues. Operations such as aggregating or reshaping a table fall under \textit{Transform (A.T)}. Finally, journalists often \textit{Export (A.E)} their data at the conclusion of the process.

\textbf{Processes}: Processes reflect our interpretations of the wrangling process. We record seven high-level descriptive categories reflecting our interpretations of the wrangling process. First, we take note of how journalists acquire data whenever such information is apparent or explicitly mentioned under \textit{Source (P.S)} of the dataset. The category \textit{Workflow (P.W)} captures user behavior as it relates to additional wrangling architecture within the environment. Although our analysis focuses on how data is wrangled, where we can confidently infer motives, we record these observations in the \textit{Cause (P.C)} branch. We record several high-level patterns for how data is transformed in \textit{Themes (P.T)}. While the scope of this project focuses on the pre-analysis activities of data journalists, we also include some observations about high-level \textit{Analysis (P.A)}. Analyzing multiple notebooks reveals a few recurring methods for the \textit{Management (P.M)}. Finally, we infer \textit{Pain Points (P.P)} encountered during the wrangling process.

\subsection{Low-Level Group Structure}

The fourteen axial codes described above comprise only the top two levels of the taxonomies. Within each of these top levels, our high- and medium-level axial codes tend to converge upon high-level wrangling tasks addressed in related work, while our lower-level axial codes and open codes illustrate salient and nuanced differences within these well-known categories. For example, \textit{Resolve Entities (A.C.b.3} refers to the common task of entity resolution, reconciling separate and distinct entries for the same real-world entity~\cite{benjelloun_swoosh_2009}. Other codes in the same group all address how users \textit{Replace (A.C.b)} values in saliently different ways, such as: \textit{Replacing NA Values (A.C.b.1)}, \textit{Editing Values (A.C.b.2)}, and \textit{Scaling Values (A.C.b.5)}. Likewise, \textit{Replace (A.C.b)} belongs to a group of codes generally concerned with \textit{Cleaning (A.C)} datasets of errors that also include \textit{Remove (A.C.a)} and \textit{Reformat (A.C.c)}.

\section{A Multi-Table Framework for Data Wrangling}
\label{sec:framework}

A key finding from the first two phases of our work is the discrepancy between journalists' frequent use of multiple tables and the single-table emphasis of most wrangling frameworks. We even found examples of journalists using multiple tables when wrangling a single data source. Many programming languages and packages support the concept of a table as a first-class object containing heterogeneous data, such as data frames in R and Pandas for Python. However, this convention is largely absent from GUI-based wrangling tools. Most interactive wrangling applications, such as Wrangler, OpenRefine, and Workbench, support only what we call a single-table wrangling context: the interface is designed around a single matrix where the table constitutes the environment, with rows and columns as the objects being wrangled within that environment. Motivated by the finding that wrangling across multiple tables is an unmet need, we present a concise multi-table framework for data wrangling. It features a small but complete set of operations that could serve as the formalism that underlies an interactive wrangling application.

\subsection{Framework Overview}

As we illustrate in Table~\ref{tbl:multi_framework}, the design space of our framework is structured by two primary dimensions. The first dimension is the type of data object, with three different types: rows, columns, and tables. While tables are collections of rows and columns, we count them as distinct entities in order to describe table-based operations that operate on rows and columns collectively. The second dimension comprises operation categories, with five top-level classes: create, delete, transform, separate, and combine.  

These five class groupings are based on the cardinality of the set of input and output objects according to three bins: zero, one, and many. Create has no inputs and one output; delete is the inverse. Transform is one-to-one. Separate has one input and many outputs, and combine is the inverse. For simplicity, we do not include many-to-zero operations as these can be described as repeated one-to-zero operations. We also restrict operations to stay within data types. For example, a transformation operation that takes one table as input could not output two columns; it would output one table.

We designed this framework to be as concise as possible, and so that all intersections between the five top-level operation classes and the three data types would be semantically meaningful. The simple create and delete operations suffice for all three data types, and the other three categories have one level of further subdivision into two or three operations each, yielding 21 operations in total.

\begin{table}[h!]
\vspace{-0.5em}
\label{tbl:multi_framework}
\begin{tabular}{ p{0.1\textwidth}|p{0.05\textwidth}|p{0.27\textwidth} }
    \textbf{Op Class} & \textbf{Sets} & \textbf{Expanded Operations} \\
    \hline
    \textit{Create} & 0:1 & T/C/R: \texttt{Create} \\
    \hline
    \textit{Delete} & 1:0 & T/C/R: \texttt{Delete} \\
    \hline
    \textit{Transform} & 1:1 & T: \texttt{Rearrange}, \texttt{Reshape}; \newline C/R: \texttt{Transform} \\
    \hline
    \textit{Separate} & 1:N & T: \texttt{Subset}, \texttt{Decompose}, \texttt{Split}; \newline C/R: \texttt{Separate} \\
    \hline
    \textit{Combine} & N:1 & T: \texttt{Extend}, \texttt{Supplement}, \texttt{Match}; \newline C: \texttt{Combine}; \newline R: \texttt{Summarize}, \texttt{Interpolate} \\
\end{tabular}
\smallskip
\caption{Multi-table framework for data wrangling. One axis of the design space is the type of data: Tables (T), Columns (C), and Rows (R).  The 5 classes of operations comprise a second axis, based on cardinalities of the input and output sets: zero, one, and many.}
\vspace{-1.5em}
\end{table}

We now describe each class of operations in the framework and discuss its connection to taxonomy classifications, database operations, and previous wrangling frameworks. 

\subsection{Create Operations}
An operation that transforms zero objects into one or more is effectively creating data objects inside the wrangling environment. We consolidate both zero-to-one and zero-to-many operations into this class because we conceptually view the latter as repeating the former.

\icon{figures/table-icons/multi-table-figures_table-0-1-define}\textbf{Create Tables}: Users in our technical observation study defined tables in three distinct ways. First, tables can be \textit{Fetched (A.I.a)} from an external source, such as a HTTP request to publicly accessible API. Second, tables can be \textit{Created (A.I.b)} directly in the wrangling environment. Third, tables can be \textit{loaded (A.I.c)} into the environment locally from a file or database residing on the user's hard drive.

\icon{figures/table-icons/multi-table-figures_column-0-1-insert} 
\textbf{Create Columns}: While new columns can be added to a table by merging another table or transforming existing columns within a table, column creation involves adding columns to tables without these sources. We observed a salient instance in our technical observation study: the code \textit{Generate Dataset Identification (A.D.c.2)} describes when a journalist defines a column of constant values, such as the year of the data or the file name. This dataset variable serves as a unique identifier for the table, and this action often occurs prior to merging tables together row-wise, \textit{Union Datasets (A.M.a)}.

\icon{figures/table-icons/multi-table-figures_rows-0-1-define} 
\textbf{Create Rows}: While row creation is not a common class of data wrangling operations, users may need to do so in order to enter missing observations from a dataset obtained by other means, which we coded as \textit{Construct Data Manually (A.I.b.1)}.

\subsection{Delete Operations}
Functions that map one or more objects to zero objects.

\icon{figures/table-icons/multi-table-figures_table-1-0-delete}
\textbf{Delete Tables}: Tables can be deleted either explicitly or implicitly. When merging multiple tables together, constituent tables may be explicitly removed after the operation to clean up the wrangling environment. In a wrangling environment where tables are first-class objects, filtering a table by rows and columns can be conceptualized as a composite task in which they \texttt{Separate} one table into two tables and implicitly \texttt{Delete} either one.

\icon{figures/table-icons/multi-table-figures_column-1-0-delete} 
\textbf{Delete Columns}: One-to-zero operations with columns deleted from the table~\cite{kandel_wrangler_2011}. Columns may be irrelevant, incomplete, or duplicate variables. Journalists in our technical observation study frequently \textit{Removed Variables (A.D.d.1)}. \textit{Merging (A.M)} datasets together may result in duplicate variables, and journalists may choose to \textit{Remove Duplicate Variables (P.P.d.4)}. Journalists also \textit{Trim Fat (P.T.d)}, removing many variables at the initial portion of wrangling. Many interactive wrangling applications~\cite{noauthor_trifacta_2012, huynh_openrefine_2012, stray_workbench_2018} support dropping multiple columns at a time. However, we omit a many-to-zero column operation category since this action is conceptually a composite of many one-to-zero column operations.

\icon{figures/table-icons/multi-table-figures_rows-1-0-delete} 
\textbf{Delete Rows}: Maps one or more rows to zero rows~\cite{kandel_wrangler_2011}. It is an essential class of wrangling operations in data preparation tasks, such as filtering. As with columns, deletion is a means of addressing irrelevant, incomplete, or duplicate observations in a dataset.

\subsection{Transform Operations}
Functions involving a one-to-one mapping of inputs and outputs.

\smallskip
\textbf{Transform Tables}: A rich class of operations involving structural changes to a table of varying complexity. On the simple end of the spectrum is \texttt{Rearrange}, or operations that transform the table without modifying the fundamental structure of the dataset, and on the complex end of the spectrum is \texttt{Reshape}, operations that modifying the fundamental schema of the dataset. 

\icon{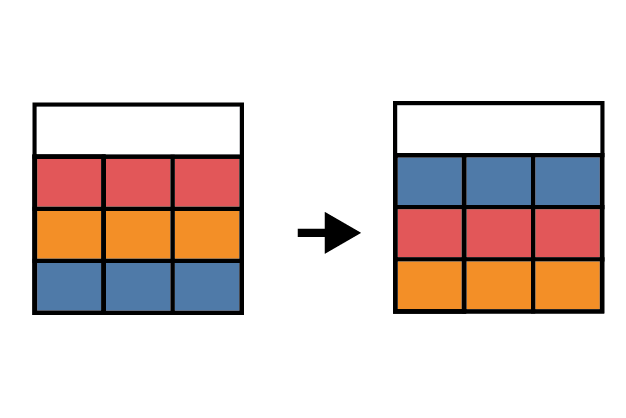}
\texttt{Rearrange} operations transform a table without fundamentally modifying the underlying table schema. Example operations in this category are \textit{sort}~\cite{kandel_wrangler_2011} and rearrange columns. While less semantically meaningful as rearranging rows, some informal conventions may facilitate presentation of a dataset, such as moving unique identifiers of an observation to the far left side of a table.

\icon{figures/table-icons/multi-table-figures_table-1-1-reshape}
\texttt{Reshape} operations change the fundamental structure of variables and observations within the dataset~\cite{kandel_wrangler_2011}. \textit{Unfold} and \textit{fold} are two examples of \texttt{reshape} operations. \textit{Fold} collapses a column into key-value sets, which can be used to restructure a table into \textit{tidy format}~\cite{wickham_tidy_2014}. \textit{Unfold} casts level values in columns of categorical variables as table columns from data values, a common method to \textit{Cross Tabulate (A.T.a.2)} data.

\icon{figures/table-icons/multi-table-figures_column-1-1-transform} 
\textbf{Transform Columns}: Map the values from one column into another column by various means, including extract, cut, and split~\cite{kandel_wrangler_2011}. Transformation can be a means to a multitude of data wrangling ends. For example, \textit{Resolving Entities (A.C.b.3)}, the process of reconciling separate and distinct entries for the same real-world entity~\cite{benjelloun_swoosh_2009}, can involve mapping a column of categorical variables with duplicate levels for the same real-world entity to a smaller, unique set of levels.

\icon{figures/table-icons/multi-table-figures_rows-1-1-edit}
\textbf{Transform Rows}: Transform one row into another because data is either incomplete or inaccurate. We observed journalists manually editing individual rows when raw data contained clerical errors, coded as \textit{Impute Missing Data (A.I.b.4)}. Errors arising from human data entry are a common issue for journalists~\cite{groskopf_quartz_2015}.

\subsection{Separate Operations}
Transform one object in the wrangling environment into many, by which we mean strictly greater than one. While our framework has parallels to operations for combining tables using database tools, these tools lack high-level support for separating tables by value.

\smallskip
\textbf{Separate Tables}: Our framework is the first to address operations in this category separating one table into many. We classify operation in the \textbf{Separate} category into three subcategories: \texttt{Subset}, \texttt{Decompose}, and \texttt{Split}.

\icon{figures/table-icons/multi-table-figures_table-1-n-subset}
\texttt{Subset} operations describe dividing a table row-wise into two tables. Filtering a table is a composite task combining \texttt{Subset} and \texttt{Delete}. After separating a table, the subset of non-matching rows is deleted from the wrangling environment.

\icon{figures/table-icons/multi-table-figures_table-1-n-decompose}
\texttt{Decompose} operations are similar to \texttt{Subset} except a single table is partitioned into any number of constituent tables based on the values and data type of a single table column. If this column represents a categorical variable, then constituent tables are divided by unique level value. In the case of boolean variables, this operation outputs two tables for true and false values. If the variable is quantitative or ordinal, the constituent table is arbitrarily divided into disjoint sets within the variable's range.

\icon{figures/table-icons/multi-table-figures_table-1-n-split} 
\texttt{Split} operations describe mapping one table to many by dividing a table column-wise. While these operations essentially produce two sets of disjoint columns, one duplicate key column always remains in order to maintain continuity between observations in the two tables.

\icon{figures/table-icons/multi-table-figures_column-1-n-split} 
\textbf{Separate Columns}: Separate a composite column into its atomic components, a necessary step for anomaly detection~\cite{raman_potters_2001}. One of the most common issues with messy data is for multiple dataset variables to be stored in a single column~\cite{wickham_tidy_2014}.

\icon{figures/table-icons/multi-table-figures_rows-1-n-separate}
\textbf{Separate Rows}: Separating one row into many can address more fundamental data parsing issues that arise in data wrangling~\cite{kandel_wrangler_2011}. OpenRefine~\cite{huynh_openrefine_2012} enables users to work with observations in multiple rows as \textit{records} in the application. This class of operations could be useful for initially parsing data stored in files with idiosyncratic structuring.

\subsection{Combine Operations}
Many-to-one transformations of objects in the wrangling environment effectively combine objects.

\smallskip
\textbf{Combine tables}: Data integration, constructing a single schema for unified access to multiple data sources~\cite{smith_multibase_1981}, is often cited as a sub-process of data wrangling~\cite{kandel_wrangler_2011,kandel_research_2011,bigelow_origraph_2019}, but details of transformations during integration are minimally addressed in existing process theories. Informed by the results of our technical observation study, we offer three subcategories of operations that combine multiple tables into one: \texttt{Extend}, \texttt{Supplement}, and \texttt{Match}.

\icon{figures/table-icons/multi-table-figures_table-n-1-extend}
\texttt{Extend} operations describe the row-wise merger of multiple tables into one table, similar to a \texttt{UNION} operator in SQL. These operations can be significantly complicated by \textit{Schema Drift (P.P.g)}, in which periodically published datasets change over time. While enterprise data analysts have been documented encountering this issue through redundant columns containing the same variable, our technical observations show that the levels of categorical variables often also change over time.

\icon{figures/table-icons/multi-table-figures_table-n-1-supplement}
\texttt{Supplement} operations incorporate data from other tables through the column-wise merger of multiple tables where tables are matched on corresponding key columns~\cite{kandel_wrangler_2011}, similar to an \texttt{OUTER JOIN} operation in SQL. There is a bijective relationship between levels in the key column of the main table and the supplementing table, distinguishing it from \texttt{Match}. One common application of supplement involves \textit{Creating Lookup Tables (P.A.m)}. Hence, using a lookup table can be a table-focused way of transforming columns.

\icon{figures/table-icons/multi-table-figures_table-n-1-match}
\texttt{Match} operations also concern the column-wise combination of many tables into one table; however, there exists an injective relationship between the matching keys of the two tables\, similar to an \textit{INNER JOIN} operation in SQL. Hence, some rows purposefully do not have corresponding matches, excluding them from the output table, also known as filter joins~\cite{wickham_dplyr_2019}. We do not specify an inverse operation for match because rows are excluded from this process by definition. To \texttt{Match} tables when the intent was to \texttt{Supplement} can cause rows to be dropped from the output table unknowingly. We code this phenomenon as a \textit{Lossy Join (P.P.d.3)}.

\icon{figures/table-icons/multi-table-figures_column-n-1-consolidate}
\textbf{Combine Columns}: Map many columns into one column. While a column containing a decomposed variable, such as an address, may need to be separated in order to perform operations on its constituent components, the data wrangler may prefer for these components to be \textit{combined} into a single column. This example is essentially a \textit{split, compute, and merge} strategy for data cleaning~\cite{wickham_split-apply-combine_2011} applied to wrangling.

\textbf{Combine Rows}: Take multiple rows as input, then output one row. We make a distinction between two combine operations within this category: \texttt{Summarize} and \texttt{Interpolate}.

\icon{figures/table-icons/multi-table-figures_rows-n-1-summarize}
\texttt{Summarize} operations combine rows, grouping observations by a categorical variable and applying an aggregate function to another variable. Aggregation effectively coarsens the granularity of the observations in a dataset. While this operation could be construed as a one-to-one transformation of tables, categorizing it as an operation upon rows better captures this coarsening effect. 

Many data analysis packages support aggregations grouped by the levels of one or more categorical variables. Aggregation alone does not completely describe this common use case. We consider this particular operation a composite operation consisting of three separate operations, similar to the Split, Apply, Combine strategy for data analysis~\cite{wickham_split-apply-combine_2011}. The same operation can be described in this framework by \texttt{Decompose}, \texttt{Summarize}, and \texttt{Extend}.

\icon{figures/table-icons/multi-table-figures_rows-n-1-interpolate}
\texttt{Interpolate} operations are another common way to combine rows. Also known as \textit{fill}~\cite{kandel_wrangler_2011}, one output row is calculated based on the values of multiple input rows. This operation can be used to address issues with missing data.

\subsection{Assessment}

\begin{figure}
    \centering
    \includegraphics[width=8cm]{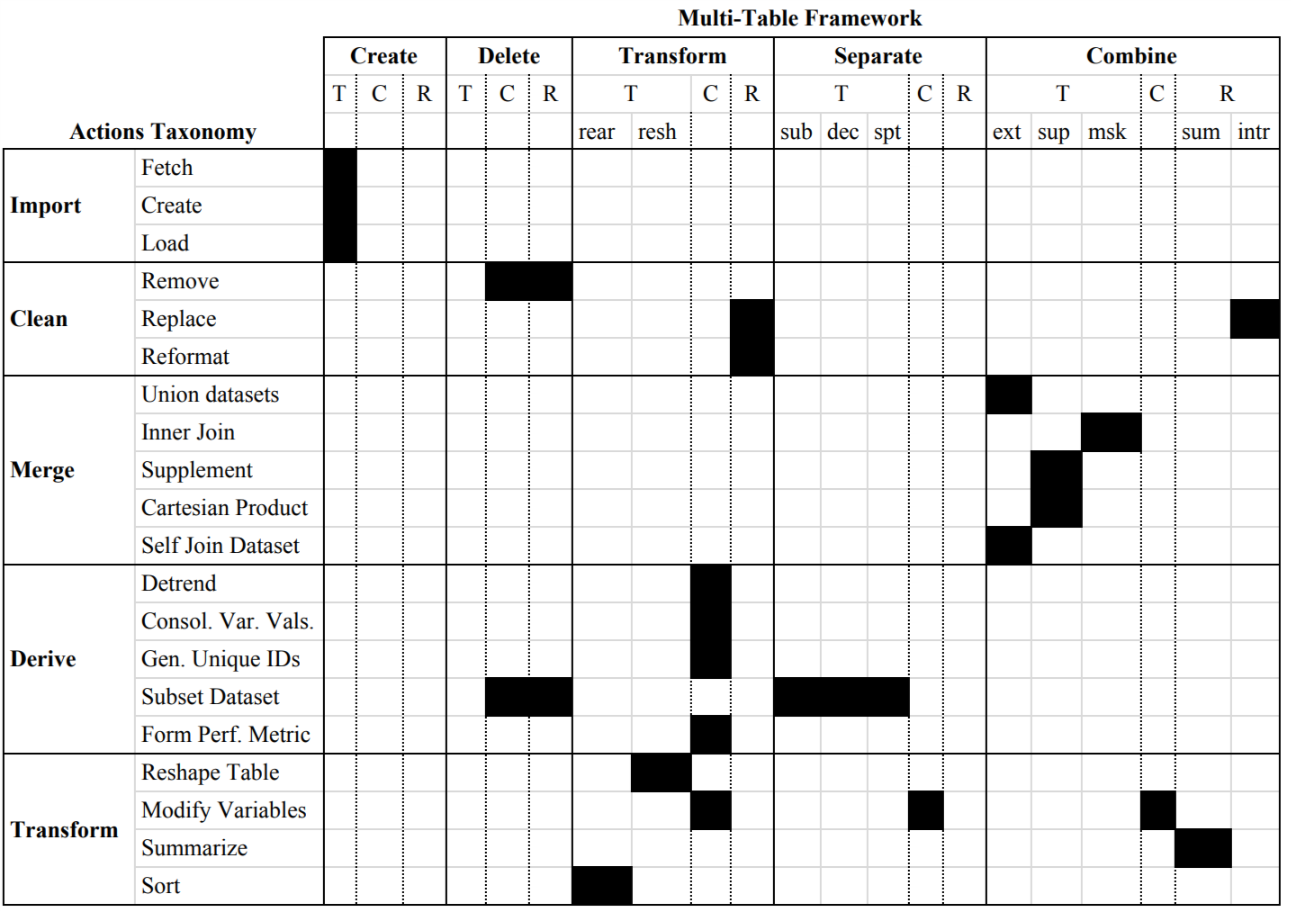}
    \caption{We cross check the descriptive power of our multi-table framework for data wrangling by comparing against the high-level axial codes in our descriptive action taxonomy. We only include Actions codes that correspond with table operations, excluding codes in the Profile branch.}
    \label{fig:cross-check}
    \vspace{-1.5em}
\end{figure}

We developed this second model for data wrangling in pursuit of greater generative power. The composability and conciseness of this 21-operation framework the 165-code set of descriptive taxonomies reasonably indicates our success. Individual operations can be arranged to create new multi-table wrangling processes and describe existing ones. For example, the popular Split, Apply, and Combine strategy of data analysis~\cite{wickham_split-apply-combine_2011} can be translated into a sequence consisting of table separation, column transformation, and table combining operations. However, our framework preserves the multi-table context of the "split" and "combine" operations and the within-table context of the "apply" operations. Figure~\ref{fig:cross-check} illustrates similar areas of descriptive overlap between these two conceptual models. We cross-checked them to ensure our multi-table framework adequately covers wrangling observations performed by journalists during our technical observation study.

\section{Critical Incidents}

This section presents two case studies taken from the corpus of 50 repos to highlight specific critical incidents. The first incident illustrates how journalists overcome challenging data wrangling tasks that cannot be accomplished with existing interactive wrangling systems. The second incident demonstrates that common mistakes when wrangling data with multiple data sources may lead to factual errors in published stories. We explain how we arrived at some of the open codes in the taxonomies.

\subsection{A Multi-Table Success Story}

\begin{figure}
    \centering
    \includegraphics[width=8cm]{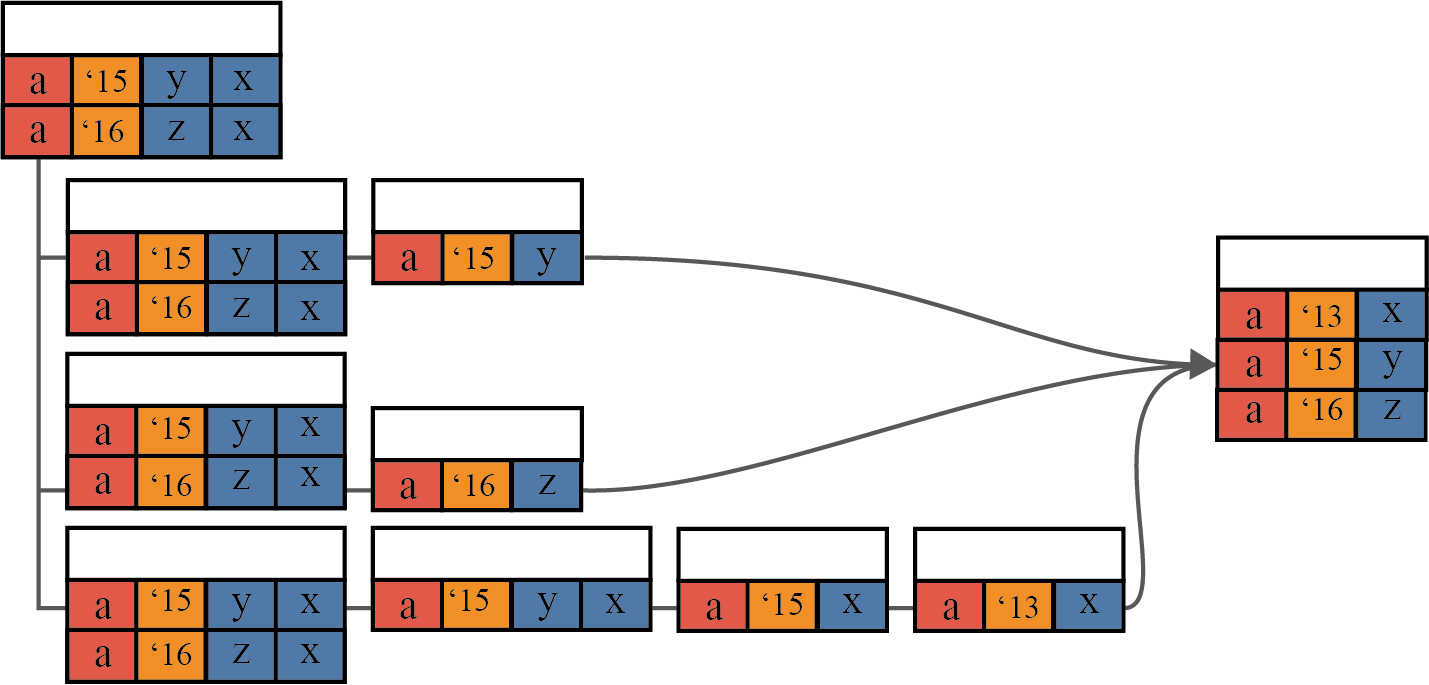}
    \caption{Journalists at the Los Angeles Times employ multiple tables to wrangle water usage data into tidy format. With water usage amounts in a separate column, common reshape operations that operate within the context of one table fail on this table.}
    \label{fig:lat-reshape}
    \vspace{-1.5em}
\end{figure}

The state of California enacted mandatory water use restrictions from 2015 to 2016, following years of record drought. \textit{Using open-government data portals (P.S.b.5)} published monthly by California’s Water Resources Control Board, reporters at the Los Angeles Times found the majority of water districts increased their usage after the state relaxed restrictions in 2016 when compared to water use in 2013, before the governor proclaimed a state of emergency due to drought.

In ``The drought eased up, and these Californians turned on the spigot'' reporters analyze this data and rank water supply agencies by a conservation-consumption score, a composite metric considering overall water savings and per-capita water use. Reporters specifically \textit{Compare Groups (P.A.b)} along a common metric to evaluate water usage in June, July, and August between 2013, 2015, and 2016.

One significant wrangling challenge involved tidying~\cite{wickham_tidy_2014} the data. Figure \ref{fig:lat-reshape} illustrates a simplified representation of the data schema at the beginning of this process, the state of intermediate representations, and the final tidy format. The dataset includes the following variables: supplier name, month and year of the water usage, and the amount of water used. This abstract dataset could easily lend itself to a three-column table. However, the actual table representation has a fourth column: the amount of water consumed in 2013. The most complicated form of messy data has different variables stored in both rows and columns, such as a cross-tabulated table~\cite{wickham_tidy_2014}. In this example, the same variable (year of recorded water production) exists in both rows and columns.

In order to tidy this dataset, the reporters implement a \textit{Divide and Conquer (P.T.a)} strategy, splitting this table into three tables, then filtering the table upon a facet. With values for 2013 in a separate table, journalists were able to apply typical wrangling operations common in single-table contexts. A new date column can be derived from the old one, unilaterally replacing any year with 2013. The column containing non-2013 values was split from the rest of the table; as was the 2013 column in the other two tables. Finally, the three constituent tables were extended together into one tidy table.

Wrangling applications that operate in a single-table context, would fail on this transformation. A logical within-table approach for this type of problem would be to \textit{Transform (A.T)} the schema via a \textit{Reshape (A.T.a)} operation. Reshaping generally captures transposing sections of the data from rows to columns, and vice versa. We observed many journalists performing this through \textit{melt} and \textit{cast} in the reshape library in R. This operation is also supported in both OpenRefine under \textit{Transpose}, and in Google Cloud Dataprep under \textit{pivot}, \textit{unpivot}, and \textit{convert columns to values}. In either approach, there would not be a straightforward way to fold 2013 values into the other columns.

\subsection{A Multi-Table Cautionary Tale}

On November 19, 2015, BuzzFeed News published the article ``Where U.S. Refugees Come From—And Go—In Charts.'' This piece presents the results of exploratory data analysis on a dataset related to refugee immigration to the United States from 2005 to 2015, with totals greater than 672,000 refugees. This data comes from the data portal for the US State Department's Refugee Processing Center. Like the other case study, the raw data is from an \textit{Open-government Data Portal (P.S.b.5)}.

Because each table represents the same phenomenon, the sum of arrivals should be equal. Journalist begins with \textit{Testing for Equality (A.P.a.2)} by summing the arrivals column of both tables and comparing the total. The religion table had four more arrivals than the destination table. No further steps were taken to address this discrepancy, coded as \textit{Tolerate Dirty Data (P.M.b.2)}. However, our own auditing of the data reveals more nuanced differences between the two tables. When comparing arrival counts by country, the destination table has one more arrival from Iran and five fewer arrivals from Iraq than the religion table, leading to a discrepancy of six arrivals between the two tables.

Although this difference is insignificant, in theory the grouped differences between arrival totals could have been arbitrarily large and the non-grouped difference could appear small. Thus, simple column sums are a problematic way to \textit{Test for Equality (A.P.a.2)} between two tables.

The majority of wrangling in this notebook involves transforming each table to make it suitable for different visualizations, \textit{Wrangle Data for Graphics (P.C.a.1)}, including total arrivals over time, arrivals from a specific country, and arrivals by destination in the US. Like the previous case study, the analysis of this notebook revolves around visually \textit{comparing groups (P.A.b)}. The final chart in this notebook compares US states by refugees admitted, normalized per 1,000 residents. The wrangling portion of this task required introducing a third table of the US state populations to the wrangling environment in order to \textit{formulate a performance metric (A.D.e)} for a fair comparison, the normalized rate of arrivals per state population. To form this derived table, the US state population and refugee arrival datasets were merged by a \textit{Outer Joining (A.M.c.1)} of tables. Our multi-table framework captures this kind of table transformation under \textit{Supplement (A.M.c)}, combining two tables that significantly increases the number of columns while rows remain unchanged.

However, the number of rows from the resulting composite table was significant in this instance. The state of Wyoming was missing from the chart when the article was first published. Later that day, the state was added to the chart and a correction was issued. Because the journalist published the code underlying this article, we know that this error resulted from a subtle issue when performing outer joins between two tables. Wyoming did not have a column in the aggregated refugee arrival table because the state did not accept any refugees between 2005 and 2015. Hence, when the state population table was outer-joined to the refugee table, Wyoming was silently dropped because the corresponding key did not exist in the other table. 

The Observations branch of the taxonomy contains many observations of \textit{Pain Points (P.P)} from our technical observation study. One frequent pain point concerned data cleaning tasks that result from merging multiple datasets, \textit{Post-merge Clean Up (P.P.d)}. This issue of silently dropping the state of Wyoming was coded as a \textit{Lossy Join (P.P.d.3)}, when data is lost after merging two tables column wise. While the implications of this error were relatively minor in the context of the whole article, this case illustrates that issues in data wrangling can have real editorial consequences for journalists.

\section{Implications for Visualization Design}
\label{sec:implications}

Visualization is an integral component of interactive wrangling, bridging the gulf of evaluation~\cite{norman_design_1990} by communicating the effects of intended operations on a dataset. Our results have design implications for visualization, including how to develop interactive interfaces, address common pain points, communicate data provenance, and recommend transformations via mixed-initiative guidance.

The clearest need is to construct an interactive visualization-based wrangling application that fully instantiates all of the operations in our multi-table framework. Although many applications support subsets, none covers them all. Combining tables is widely supported, similar to \textit{JOIN} operations in SQL, but no application fully supports operations for separation, which would help resolve the common problem of one table with multiple types of observations~\cite{wickham_tidy_2014}. Our multi-table framework specifies an underlying formalism, but interaction design is a remaining challenge. An approach similar in spirit to Polaris~\cite{stolte_polaris_2002} would be suitable, with a bidirectional mapping between a formalism and actions carried out through drag-and-drop visual interface.

Through code comments authored by journalists and inferences based on our own experience wrangling data, we observed many pain points that are amenable to a visualization-based solution. Visualization could aid journalists in detecting \textit{Schema Drift (P.P.g)} when wrangling perennially published datasets, which includes changes in dataset variable names and values over time. Combining tables may introduce errors in the wrangling process, which wranglers further address as \textit{Post-Merge Clean Up (P.P.d)}. In our second case study, a political map of the United States could have alerted the wrangler that their exported dataset was missing a state. In general, visualization paired with semantic and accurately inferred variable types can alert wranglers to missing data in common geographic and temporal variables.

Data provenance is more complicated in multi-table wrangling processes, and visualizations of these processes as data-flow diagrams can concisely depict these complex flow networks. A major visualization challenge for the few interactive wrangling applications that incorporate this idiom~\cite{noauthor_trifacta_2012, noauthor_tableau_2019} and any future multi-table wrangling tool, is designing automatic layout algorithms to draw comprehensible data-flow diagrams for workflows as complicated as we observed among journalists. These workflows possess many qualities that contribute to visual complexity: cyclic processes, multiple sinks, dozens of sources, and even more interior nodes.

Finally, a mixed-initiative system could facilitate and expedite wrangling through automatically generated recommendations, where the user could visually preview data transformations before interactively selecting which ones are desirable. Although Wrangler~\cite{kandel_wrangler_2011} provides this capability for single-table wrangling, multi-table support would require not only the improved provenance diagrams mentioned above, but also the ability to show effects across multiple tables, possibly with semantic zooming or other multi-scale approaches~\cite{lam_guide_2010}.

\section{Discussion}

Many interactive wrangling applications specify a design space of supported transformations informed by data transformation languages~\cite{kandel_wrangler_2011} or personal experience in wrangling data~\cite{bigelow_origraph_2019}. To the best of our knowledge, no one has generated a design space of data transformations grounded in observational data gathered from users wrangling data in the wild, with the full flexibility of programming through script-based languages. We conjectured that these programmatic approaches to wrangling might be more expressive than those supported by interactive wrangling applications, and indeed we found these differences. We were especially interested in patterns of behavior where users appear to be exerting a lot of effort to accomplish a relatively simple task. Such discrepancies between the level of effort and the simplicity of the task can signal deficiencies in a particular model of wrangling.

A better understanding of journalistic data wrangling holds considerable promise for positive social impact. Journalism provides a public good, especially through investigative and public affairs reporting that uses data to hold corporations, public institutions, and elected officials accountable~\cite{hamilton_democracys_2018}. Better support for the time-consuming, error-prone practices of data wrangling could lead to tools that better support journalists in conducting this socially beneficial work. Even as data-driven journalism grows more important in many newsrooms, layoffs have diminished their on-the-ground reporting capacity: staff at US newspapers declined 47 percent from 2008 to 2018~\cite{grieco_us_2019}, but better data analysis tools can help news organizations meet the unfortunate demand to do more with less. Visual analytics researchers need to target journalists as an often-overlooked group who specifically needs better tools. When journalists are lumped in with non-technical users, two problems emerge. First, we ignore the growing number of computational journalists who are quite technical and use data in their daily work. Second, tools built for the general user may not be as effective for journalists if they do not match the domain need.

While we thoroughly searched for relevant repos, our data collection method is still inherently biased towards completed projects that yielded newsworthy findings. Thus, we do not claim to offer complete coverage of wrangling activities in journalism. In future work, we plan to explore instances of unsuccessful wrangling through an interview study.

Although excluding failures is a limitation, our corpus of coded notebooks still contains telling instances of unsuccessful wrangling. Success in data wrangling does not solely depend on coercing data into an acceptable state of utility. Expending a reasonable amount of time and effort in relation to the complexity of the task is also an essential criterion for success. Both critical incidents illustrate varying degrees of failure in data wrangling despite producing data for further analysis. Our second critical incident illustrates how data in unacceptable state resulted in real editorial consequences. Journalists in our first example were able to wrangle raw data into a useful state. However, constructing seven intermediate data products is an unreasonable amount of effort for tidying a table. A wrangling process that comprises up to 80-90\% of an analyst's time~\cite{dasu_exploratory_2003,muller_how_2019} should not be considered successful. Usability metrics, such as time to completion, ought to be an essential measure of success when evaluating interactive wrangling applications, but these have only been included in a few previous system evaluations~\cite{kandel_wrangler_2011}.

We observe that the raw data in many of our repos is significantly better structured than previous work examples, so some amount of wrangling may have been performed prior to the journalists' actions. Our observational data may reflect a \textit{last-mile} problem in data wrangling, where raw data comes in a structured format, but requires further tweaking to match the user's mental model. How journalists operate on data in the last mile may correlate with different types of data journalism story. The \textit{Analysis (P.A)} branch of our Process Taxonomy shares some salient similarities with taxonomies of data journalism stories~\cite{veglis_towards_2017}, including \textit{Compare Groups (P.A.b)}, \textit{Show Trend Over Time (P.A.d)}, and \textit{Identify Extreme Values (P.A.c)}. For example, \textit{Comparing Groups} often involves several critical operations in the Actions branch: \textit{Generating Unique Identifiers (A.D.c)} for each group, \textit{Formulating a Performance Metric (A.D.e)} by which to fairly compare groups, and \textit{Summarizing (A.T.c)} this derived data. An interactive wrangling application designed with awareness of common, high-level analysis goals could further expedite the wrangling process for journalists.

\section{Conclusion}

Our paper answers three questions. First, how do journalists wrangle their data? By employing a technical observation study, we produce two richly detailed descriptive taxonomies of data wrangling in computational journalism to answer this question. Although we find that journalists perform many familiar wrangling tasks that are well supported by previous wrangling frameworks, they need far more support wrangling involving multiple tables than previous work acknowledged. Second, do journalist behaviors and needs match up with existing literature on data wrangling? We find that previous process frameworks for wrangling do not incorporate the concept of a table as a first-class object in the environment. Third, how can we re-characterize wrangling to match the needs of journalists? We present a concise framework that describes required operations to support multi-table wrangling, with actionable generative power that could support future interactive tools.

\acknowledgments{We thank Jürgen Bernard, Anamaria Crisan, Madison Elliott, Zipeng Liu, Joanna McGrenere, Francis Nguyen, Michael Oppermann, and Ben Shneiderman for their formative feedback. This work was supported by NSERC CREATE 386138851 and Discovery RGPIN-2014-06309.}

\bibliographystyle{abbrv}

\bibliography{table-scraps}
\end{document}